\documentclass[%
preprint, 
superscriptaddress,
amsmath,
amssymb,
aps, 
pra,
]{revtex4-2}

\usepackage{graphicx}
\usepackage{dcolumn}
\usepackage{bm}
\usepackage{hyperref}
\usepackage{color}

\begin{document}


\title{Products of displaced Laguerre--Gaussian beams}%

\author{G. Mellado-Villase\~nor}
\affiliation{Facultad de Ciencias, Universidad Nacional Autónoma de México, Ciudad de México, 04510, Mexico}
\email{gmelladov@ciencias.unam.mx}

\author{B. M. Rodr\'iguez-Lara}%
\affiliation{Universidad Polit\'ecnica Metropolitana de Hidalgo, Tolcayuca, Hidalgo 43860, Mexico.}%
\email{blas.rodriguez@gmail.com}


\date{\today}

\begin{abstract}
We study the free-space propagation of products of displaced Laguerre--Gaussian beams.
Each displaced factor admits an exact representation as a superposition of standard Laguerre--Gaussian beam modes through the optical analog of displaced number states.
We reduce the resulting product to a single modal expansion with closed-form weight coefficients and explicit azimuthal selection rules.
Working in the reference frame defined by the centroid of the transverse displacements, we evaluate the net orbital angular momentum directly from the modal weights, which provides a criterion to predict transverse rotation.
We identify three propagation regimes: no transverse rotation for zero net orbital angular momentum, rigid rotation for products of identical factors with zero radial index, and nonrigid rotation with intensity redistribution otherwise.
Our framework enables engineering structured light beams whose transverse rotation encodes propagation information, relevant to depth-sensitive optical microscopy, imaging, and tracking. 
\end{abstract}

\maketitle
\newpage 

\section{Introduction}
\label{sec:Sec1}

Laguerre--Gaussian beams (LGBs), solutions to the paraxial wave equation \cite{Siegman1986}, play a central role in optical trapping \cite{Dholakia2011,Grier2003} and optical communications \cite{Wang2012} due to their structured intensity profile and helical phase. 
The radial index $p$ sets the number of concentric nodes in the intensity distribution, and the azimuthal index $\ell$ determines the number of $2\pi$ phase windings around the origin \cite{Allen1992}. 
As an orthogonal basis, LGBs support linear superpositions that enable beam shaping and mode synthesis \cite{Aguirre2025}.

Products of LGBs arise in nonlinear optical processes when the input beams are LGBs.
In second-harmonic generation, two identical LGBs at the fundamental frequency generate a beam at twice the frequency, preserving orbital angular momentum and retaining the radial intensity profile only for modes with $p=0$ \cite{Courtial1997}.
In sum-frequency generation, two distinct LGBs combine according to modal constraints that correlate their radial and azimuthal indices \cite{Wu2020}.
Nonlinear interactions between structured spatial modes have been used to control modal transfer and improve structured-output fidelity \cite{Singh2025p562028,AguilarCardoso2025p043541}.
Products of three or more LGBs will appear in higher order nonlinear interactions.

The relevance of these products extends to linear propagation. 
Squared LGBs display self-similar propagation and pattern revivals in free space \cite{Kozlova2023,Zhong2021}, and their Fourier invariance enables compact beam transformations \cite{Kotlyar2022}. 
Diffraction-based multiplication provides a linear route to generate structured arrays and improve vortex-beam purity \cite{Samadzadeh2024}. 
Talbot-type configurations produce two-dimensional LGB lattices with non-zero radial indices, supporting applications in optical manipulation and communications \cite{Amiri2025}. 
These examples show that products of LGBs produce structured fields with propagation dynamics distinct from those of their individual factors.

This raises a natural question: what additional structure appears when the factors are displaced before forming the product.  
Displacements introduce relative phases and intensity gradients, suggesting new transverse dynamics.
To build intuition, we outline a heuristic picture of how these products evolve in free space.
We begin by isolating one factor and treating the remaining ones as an effective phase landscape. 
For short propagation distances, the gradient of this landscape determines the local transverse shift of the isolated factor. 
For example, when a propagation-invariant beam multiplies a plane wave, the isolated field shifts in the direction set by the wave vector encoded in the plane-wave phase.  
If the effective phase carries a non-zero winding, the resulting transverse shift acquires an angular component that induces rotation of the isolated field.  
Applying this picture to each factor suggests their correlated rotation during propagation.

Fields that rotate in a controlled manner during propagation provide an axial reference for three-dimensional imaging, where the rotation angle encodes depth with high sensitivity \cite{Greengard2006,Pavani2008,Wang2017}. 
Rotating point-spread functions implement this idea experimentally and enable precise axial tracking in biological \cite{Bonin2022,Zhang2025} and even space \cite{Wang2019} settings. 
These results illustrate that controlled rotation is an operational degree of freedom in free-space propagation and motivate us to examine how these beams encode rotation through their modal content and displacement.
Related dynamics of phase singularities in three-dimensional structured fields has also been explored in the context of optical knots and their robustness under perturbations \cite{Larocque2020p,Pires2025p}.

To move beyond our heuristic picture, we derive an analytical formulation for the free-space propagation of products of displaced Laguerre--Gaussian beams (pdLGBs).
Using an optical analogy with quantum displaced number states \cite{MoralesRodriguez2024a,MoralesRodriguez2024b}, we express each displaced factor as a superposition of standard LGBs centered at the origin of the transverse plane and rewrite their product as a single superposition. 
We obtain explicit complex-valued coefficients for this expansion that encode displacements and mode overlaps.
This decomposition enables direct calculation of orbital angular momentum and reduces propagation to tracking independent weighted modes. 
All components in the superposition share a common spherical phase and radial scaling set by the Gaussian beam waist, while acquiring distinct Gouy phases proportional to their total modal numbers. 
By absorbing the Gouy phase into the displacement coefficients, we reveal transverse rotations of each component governed by the net OAM of the original field. 
Our framework supports beam engineering in linear homogeneous media and provides analytical control of three-dimensional structure in vortex-beam products.

\section{Results}
\label{sec:Sec2}

Exact paraxial propagation follows from the Fresnel integral \cite{Goodman2005}
\begin{align}
    \begin{aligned}
        \Psi(\rho,\theta,z) = \frac{e^{ikz}}{i\lambda z}
        \int_{0}^{2\pi} d\theta' \int_{0}^{\infty} d\rho'\, \rho'\, \Psi(\rho',\theta',0)\, e^{\frac{i k}{2z}\!\left[\rho^{2} + \rho'^{2} - 2\rho\rho' \cos(\theta - \theta')\right]},
    \end{aligned}
\end{align}
where $(\rho,\theta)$ and $(\rho',\theta')$ denote transverse polar coordinates at the observation and input planes.  
Although exact, this representation offers limited analytical intuition beyond a few special initial configurations for which the integral can be evaluated in closed form.

To obtain a closed-form analytical representation and physical insight, 
we adopt a modal formulation that is mathematically equivalent to Fresnel propagation but analytically tractable.
We reformulate paraxial propagation without pursuing a direct evaluation of the Fresnel integral, which is not analytically accessible in general form, using the modal structure of Laguerre--Gaussian beams, together with the optical analog of displaced number states.
Each pdLGB reduces to a weighted sum of standard LGBs with known propagation.
All components in the superposition share the same spherical phase and radial scaling, while their distinct Gouy phases generate transverse rotation.
Our modal formulation provides direct analytical access to propagation behavior that is not transparent at the level of the Fresnel integral for products of displaced Laguerre--Gaussian beams.

\subsection{Laguerre--Gaussian beams}

We define scalar LGBs \cite{Aguirre2025} as
\begin{align}
    \Psi_{p,\ell}(\rho,\theta,z) = \frac{\sqrt{2}}{w(z)}\, e^{ - \frac{i k \rho^{2}}{2 R(z)}}\, e^{i \left( 2p + \vert \ell \vert + 1 \right) \varphi(z)}\, \psi_{p,\ell}(r,\theta),
\end{align}
with units of $\mathrm{length}^{-1}$ in cylindrical coordinates $(\rho, \theta, z)$.
The standard Laguerre-Gauss mode, 
\begin{align}
    \psi_{p,\ell}(r,\theta) =&~ (-1)^p \sqrt{\frac{p!}{\pi (p + \vert \ell \vert)!}}\, r^{\vert \ell \vert} e^{-r^2}\, {\text L}_p^{\vert \ell \vert}(r^2)\, e^{i \ell \theta},
\end{align}
with dimensionless radial coordinate $r = \sqrt{2} \rho / w(z)$ and generalized Laguerre polynomial ${\text L}_p^{\vert \ell \vert}(\cdot)$, defines the transverse profile. 
The modal numbers $p = 0, 1, 2, \ldots$ and $\ell = 0, \pm 1, \pm 2, \ldots$ characterize the radial and azimuthal node counts \cite{Karimi2014,Plick2015}.
We use the standard expressions for the beam waist $w(z) = w_0 \sqrt{1 + (z / z_R)^2}$, the radius of curvature $R(z) = z \left[ 1 + (z_R / z)^2 \right]$, and the Gouy phase $\varphi(z) = \tan^{-1}(z / z_R)$, with $w_0$ being the beam waist at the input plane $z = 0$ and $z_R = \pi w_0^2 / \lambda$ being the Rayleigh range \cite{Siegman1986}.
These beams form an orthonormal basis,
\begin{align}
    \begin{aligned}
         \int_{0}^{2\pi} d\theta \int_{0}^{\infty}  d\rho \; \rho \,  \Psi^{\ast}_{p,\ell}(\rho,\theta,z)\,  \Psi_{p^{\prime},\ell^{\prime}}(\rho,\theta,z) = \delta_{p,p'} \delta_{\ell,\ell'}
    \end{aligned}
\end{align}
at each transverse plane $z$ and carry an OAM proportional to the azimuthal modal number along the propagation direction,
\begin{align}
    \begin{aligned}
        \langle \hat{L}_{z} \rangle =&~  \int_{0}^{2\pi} d\theta \int_{0}^{\infty} d\rho \; \rho \, \Psi_{p,\ell}^{\ast}(\rho,\theta,z)\, \left[ - i \hbar \partial_{\theta} \right]\, \Psi_{p,\ell}(\rho,\theta,z), \\
        =&~ \hbar \ell.
    \end{aligned}
\end{align}

\subsection{Displaced Laguerre--Gaussian beams}

We consider paraxial scalar light beams constructed as the product of $n$ LGBs
centered at positions $(x_{j},y_{j})$ in the input plane $z=0$,
\begin{align}
    \Psi(\rho,\theta,0)
    =&~ \frac{1}{\sqrt{\mathcal{N}}}
    \prod_{j=1}^{n}
    \Psi_{p_{j},\ell_{j}}(x-x_{j},y-y_{j},0),
    \label{eq:PDLGB}
\end{align}
with normalization constant $\mathcal{N}$.  
Our goal is to build intuition for their structure under propagation by using a modal decomposition in terms of standard LGBs centered at the origin of the transverse plane.

We use the optical analog of displaced number states \cite{MoralesRodriguez2024a,MoralesRodriguez2024b} to obtain an exact modal expansion of each displaced factor in the product in terms of standard LGBs,
\begin{align}
    \Psi_{p_{j},\ell_{j}}(x-x_{j},y-y_{j},z)
    = \sum_{u_{j},v_{j}=0}^{\infty}
    c_{u_{j},v_{j}}
    \Psi_{q_{j},l_{j}}(\rho,\theta,z),
    \label{eq:expSLGB}
\end{align}
with modal indices
\begin{align}
    \begin{aligned}
        q_{j} =&~ \min(u_{j},v_{j}), \\
        l_{j} =&~ u_{j}-v_{j}.
    \end{aligned}
\end{align}

Our analogy provides the complex weight coefficients
\begin{align}
    \begin{aligned}
        c_{u_{j},v_{j}}
        =\,&
        \frac{(-1)^{m_{j}+n_{j}}
        \alpha_{j+}^{\,u_{j}-m_{j}}
        \alpha_{j-}^{\,v_{j}-n_{j}}}
        {\sqrt{u_{j}!\,v_{j}!\,m_{j}!\,n_{j}!}}
        \, e^{-\frac{1}{2}(\vert\alpha_{j+}\vert^{2}
        +\vert\alpha_{j-}\vert^{2})}
        \times \\
        &\times
        \mathrm{U}\!\left(-m_{j},-m_{j}+u_{j}+1,\vert\alpha_{j+}\vert^{2}\right)
        \mathrm{U}\!\left(-n_{j},-n_{j}+v_{j}+1,\vert\alpha_{j-}\vert^{2}\right),
    \end{aligned}
    \label{eq:DispCoeff}
\end{align}
where the confluent hypergeometric functions $\mathrm{U}(a,b,z)$ depend on the modal indices,
\begin{align}
    \begin{aligned}
        m_{j} =&~ p_{j}+\max(0,\ell_{j}), \\
        n_{j} =&~ p_{j}+\max(0,-\ell_{j}),
    \end{aligned}
\end{align}
and on the complex displacements,
\begin{align}
    \alpha_{j\pm}
    = \frac{1}{\sqrt{2}w_{0}}
    \sqrt{x_{j}^{2}+y_{j}^{2}}\,
    e^{\mp i\arctan(y_{j}/x_{j})},
\end{align}
which encode the transverse position of each factor.  
These coefficients satisfy
$\sum_{u_{j},v_{j}=0}^{\infty}\vert c_{u_{j},v_{j}}\vert^{2}=1$,
ensuring normalization.

\begin{figure}[ht]
    \centering
    \includegraphics[width =  0.75 \linewidth]{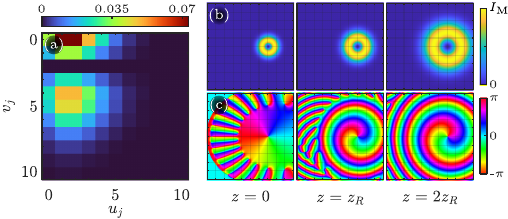}
    \caption{(a) $\vert c_{u_j,v_j} \vert^{2}$ for a displaced LGB with $ \{ p_j,\ell_j,(x_j,y_j) \} = \{0,1,(2w_{0},0) \} $ for $u_{j}, v_{j} \in \left\{ 0, 1, 2, \ldots, 10\right\}$. 
    (b) Intensity and (c) phase distributions at $z \in \left\{  0, 1, 2 \right\} z_{R}$ in the range $x, y \in [-5, 5] w_{0}$, with $I_{\mathrm{M}}$ being the maximum intensity value.
    Truncation artifacts are visible in the phase distribution at $z = 0$ and $z = z_R$.}
    \label{fig:Fig1}
\end{figure}

Displaced LGBs constructed in this manner propagate without shifting their transverse centers $(x_{j},y_{j})$, preserving their shape up to radial scaling.
Figure~\ref{fig:Fig1} shows an example with $\{p_{j},\ell_{j},(x_{j},y_{j})\}=\{0,1,(2w_{0},0)\}$.
We use the squared magnitudes $\vert c_{u_{j},v_{j}}\vert^{2}$ [Fig.~\ref{fig:Fig1}(a)] to truncate the superposition in Eq.~\eqref{eq:expSLGB}.
Figures~\ref{fig:Fig1}(b) and \ref{fig:Fig1}(c) show intensity and phase distributions at $z\in\{0,1,2\}\,z_{R}$.
Truncation introduces small artifacts visible in the phase at $z=0$ and $z=z_{R}$.
The spiral vortex structure follows from the spherical phase factor $e^{i k\rho^{2}/[2R(z)]}$.

\subsection{Products of displaced Laguerre--Gaussian beams}

We rewrite the pdLGBs at the input plane $z=0$ as
\begin{align}
    \Psi(\rho, \theta, 0) = \frac{1}{\sqrt{\mathcal{N}}} \prod_{j=1}^{n} \sum_{u_j, v_j=0}^{\infty} c_{u_j, v_j} \Psi_{q_{j}, l_{j}}(\rho, \theta, 0),
\end{align}
using their modal expansion.
Collecting terms gives a linear superposition of standard LGBs, 
\begin{align}
    \Psi(\rho, \theta, 0) = \sum_{P=0}^{\infty} \sum_{L=-\infty}^{\infty} C_{P,L} \Psi_{P,L}(\rho, \theta, 0),
\end{align}
with expansion coefficients,
\begin{align}
    \begin{aligned}
    C_{P,L} =&~  \int_{0}^{2\pi} \mathrm{d}\theta \, \int_{0}^{\infty} \mathrm{d}\rho \; \rho \, \Psi_{P,L}^{*}(\rho, \theta, 0) \Psi(\rho, \theta, 0), \\
    =&~ \frac{1}{\sqrt{\mathcal{N}}} \sum_{u_{1},v_{1} = 0}^{\infty} \ldots \sum_{u_{n},v_{n} = 0}^{\infty} \left( \prod_{j=1}^{n} c_{u_{j},v_{j}} \right) \mathcal{G}^{(n)},
    \end{aligned}
\end{align}
in terms of the mode-overlap integral, 
\begin{align}
    \begin{aligned}
        \mathcal{G}^{(n)} =&~ \int_{0}^{2\pi} \mathrm{d}\theta \, \int_{0}^{\infty} \mathrm{d}\rho \; \rho \, \Psi_{P,L}^{*}(\rho, \theta, 0) \left( \prod_{j=1}^{n} \Psi_{q_{j},l_{j}}(\rho, \theta, 0) \right), \\
        =&~ \left( \frac{\sqrt{2}}{\sqrt{\pi} \, w_{0}} \right)^{n-1}  (-1)^{P + q} \sqrt{  \frac{  \, P!}{ (P+\vert L \vert)!}} \,   \delta_{L,l} \left( \prod_{j=1}^{n}  \sqrt{\frac{q_{j}!}{ (q_{j}+\vert l_{j} \vert)!}} \right) 
        \\
        &~ \times \sum_{k=0}^{P} \sum_{m_{1}=0}^{q_{1}} \ldots \sum_{m_{n}=0}^{q_{n}} (-1)^{k+m}   \frac{1}{k!} \binom{P + \vert L \vert}{P - k} \left[ \prod_{j=1}^{n} \frac{1}{m_{j}!} \binom{ q_{j} + \vert l_{j} \vert}{ q_{j} - m_{j}} \right] \left( \frac{n+1}{2} \right)^{-\nu} \Gamma(\nu) , 
    \end{aligned} 
\end{align}
where we introduce the following shorthand notation, 
\begin{align}    
    \begin{aligned}
        q = \sum_{j=1}^{n} q_{j}, \quad
        l = \sum_{j=1}^{n} l_{j}, \quad
        \tilde{l} = \sum_{j=1}^{n} \vert l_{j} \vert, \quad
        m =\sum_{j=1}^{n} m_{j}, \quad
        \nu = \frac{1}{2} \left( \vert L \vert + \tilde{l} \right) + k + m + 1.
    \end{aligned}
\end{align}
We obtain this overlap using the Laguerre polynomial expansion \cite{Olver2010}, 
$\mathrm{L}_{a}^{b}(x) = \sum_{j=0}^{a} (-1)^{j} \binom{a + b}{a - j} \frac{x^{j}}{j!} $,
and the gamma function integral \cite{Olver2010}, $ \int_{0}^{\infty} x^{s-1} e^{-\lambda x} \, \mathrm{d}x = \lambda^{-s} \Gamma(s) $. 
The constant $\mathcal{N}$ ensures normalization, $ \sum_{P=0}^{\infty} \sum_{L=-\infty}^{\infty} \vert C_{P,L} \vert^{2} = 1 $. 

Our analytical modal decomposition at the input plane provides the foundation for interpreting the free-space propagation of the pdLGBs.

\subsection{Propagation and OAM}

The propagated pdLGB inherits its structure directly from the modal decomposition,
\begin{align}
    \Psi(\rho,\theta,z) = \sum_{P,L} C_{P,L}\, \Psi_{P,L}(\rho,\theta,z),
\end{align}
so that propagation reduces to tracking the evolution of the standard LGB basis,
\begin{align}
    \begin{aligned}
        \Psi(\rho,\theta,z)
        =&~ \frac{\sqrt{2}}{w(z)}\,
            e^{- \frac{i k\rho^{2}}{2R(z)}}
            \sum_{P,L}
            C_{P,L}\,
            e^{i(2P+\vert L \vert+1)\varphi(z)}\,
            \psi_{P,L}(r,\theta),
    \end{aligned}
\end{align}
which makes explicit the three mechanisms that determine the beam structure under propagation.

First, all modal components share the same quadratic phase factor $e^{-\frac{i k\rho^{2}}{2R(z)}}$, set by the radius of curvature $R(z)$ of the Gaussian envelope.
This common factor shapes the wavefront and produces the spiral phase patterns observed under propagation.

Second, the radial coordinate rescales as $r=\sqrt{2}\rho/w(z)$, so all components of the expansion undergo the same transverse scaling as the beam evolves.

Third, each term acquires a Gouy phase factor $e^{ i ( 2P + \vert L \vert + 1 ) \varphi(z)}$.
The Kronecker delta $\delta_{L,l}$ in the mode–overlap integral enforces a selection rule in the expansion, so that each term propagates with a Gouy phase consistent with its associated azimuthal index $ L = \sum_{j=1}^{n} l_{j}$.
This phase accumulation can be decomposed and absorbed into the corresponding displacement coefficients $c_{u_{j},v_{j}}$, inducing rotation.

Together, these three mechanisms determine how the modal content of a pdLGB evolves under free-space propagation.
Their combined effect is clear when viewed in terms of the transverse geometry of the displaced factors.
Each displaced LGB determines a point in the transverse plane through its center $(x_{i},y_{i})$, and the centroid of these points, $(x_{c},y_{c}) = \frac{1}{n}\sum_{i=1}^{n}(x_{i},y_{i})$, provides a natural reference frame for the propagated field.
Shifting the origin to $(x_{c},y_{c})$ centers the description on the axis about which the pdLGB acquires OAM under propagation.
In this frame, the net OAM,
\begin{align}
    \langle \hat{L}_{z} \rangle
    = \hbar \sum_{L=-\infty}^{\infty}
        \left( \sum_{P=0}^{\infty} \vert C_{P,L} \vert^{2} \right) L,
\end{align}
sets the transverse rotation of the field.
If $\langle \hat{L}_{z} \rangle = 0$, the pdLGB does not rotate.
For identical factors with radial and azimuthal indices $\{q_{j},l_{j}\}=\{0,\ell\}$, all displaced coefficients acquire the same weight factor and Gouy phase.
The pdLGB then evolves as a rigid transverse structure, rotating from $0$ at the input plane $z=0$ to $\pi$ at the Fourier plane $z=\infty$.
For $q_{j}>0$, even when all factors share identical indices $\{q_{j},l_{j}\}=\{p,\ell\}$, the modal-overlap integral involves multiple radial contributions providing different weight factors.
The pdLGB rotates with a redistribution of transverse intensity.

Our framework isolates the induced rotation in the centroid frame and connects the transverse propagation dynamics of pdLGBs to the Gouy phases and relative weights of their modal expansion.

\section{Discussion}
\label{sec:Sec3}
We illustrate the free-space propagation behavior of pdLGBs distinguishing between no rotation, rigid rotation, and nonrigid rotation.
Unless otherwise stated, the initial beam factors are radially displaced by half a beam waist, $\sqrt{x_{j}^{2} + y_{j}^{2}} = w_{0}/2$, at the input plane $z=0$, and the modal expansion is truncated to radial indices $P \in [ 0, 10 ]$ and azimuthal indices $L \in [ -5, 5 ]$, as guided by the modal weights $\vert C_{P,L} \vert^{2}$ shown in panels (a) of each figure.
Panels (b) and (c) show intensity and phase distributions at $z \in \{ 0, 1, 2 \} z_{R}$.
In all cases, our analytical results are in good agreement with numerical Fresnel propagation.

\subsection{No rotation}

\begin{figure}[ht]
    \centering
    \includegraphics[width =  0.75 \linewidth]{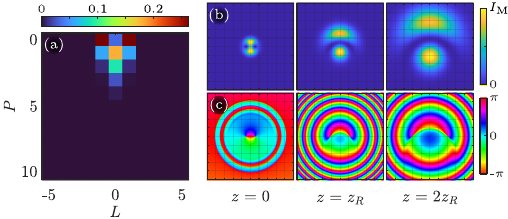}
    \caption{
    (a) Modal weights $\vert C_{P,L} \vert^{2}$ for the pdLGB formed by two displaced LGBs located at $\{ (x_{1},y_{1}), (x_{2},y_{2}) \} = \{(w_{0}/2, 0), (-w_{0}/2, 0)\}$ with opposite azimuthal charges, $\{ p_{1}, \ell_{1} \} = \{ 0, 1 \}$ and $\{ p_{2}, \ell_{2} \} = \{ 0, -1 \}$, yielding zero net OAM.
    (b) Intensity and (c) phase distributions at $z \in \{ 0, 1, 2 \} z_{R}$ in the range $x, y \in [ -5, 5 ] w_{0}$, with $I_{\mathrm{M}}$ the maximum intensity value.
    Truncation artifacts are visible in the phase distribution at $z=0$.
    }
\label{fig:Fig2}
\end{figure}

We first consider pdLGBs with vanishing net OAM, $\langle \hat{L}_{z} \rangle = 0$.
The simplest construction pairs displaced factors with opposite OAM, $\ell_{j}=-\ell_{j+1}$, arranged symmetrically with respect to the origin.
This symmetry produces modal weights $\vert C_{P,L}\vert^{2}$ that are symmetric about $L=0$.

\begin{figure}[ht]
    \centering
    \includegraphics[width =  0.75 \linewidth]{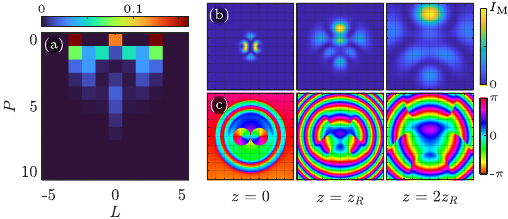}
    \caption{
Same as Fig.~\ref{fig:Fig2} for two displaced LGBs with opposite azimuthal indices and identical higher radial order, $\{ p_{1}, \ell_{1} \} = \{ 1, 1 \}$ and $\{ p_{2}, \ell_{2} \} = \{ 1, -1 \}$, located at $\{ (x_{1},y_{1}), (x_{2},y_{2}) \} = \{(w_{0}, 0), (-w_{0}, 0)\}$.
}
    \label{fig:Fig3}
\end{figure}

Figures~\ref{fig:Fig2} and \ref{fig:Fig3} show pdLGBs formed by a pair of displaced LGBs with opposite OAM, $\{ p_{1}, \ell_{1} \} = \{ p, 1 \}$ and $\{ p_{2}, \ell_{2} \} = \{ p, -1 \}$, for $p=0$ and $p=1$, respectively.
In both cases, the modal weights are symmetric about $L=0$ as expected [Figs.~\ref{fig:Fig2}(a) and \ref{fig:Fig3}(a)].
The intensity and phase distributions [Figs.~\ref{fig:Fig2}(b) and \ref{fig:Fig2}(c) and Figs.~\ref{fig:Fig3}(b) and \ref{fig:Fig3}(c)] show no transverse rotation.
For $p=1$, the additional radial structure enhances the redistribution of intensity during propagation, while the overall transverse shape and orientation remain fixed.

\subsection{Rigid rotation}
\begin{figure}[ht]
    \centering
    \includegraphics[width =  0.75 \linewidth]{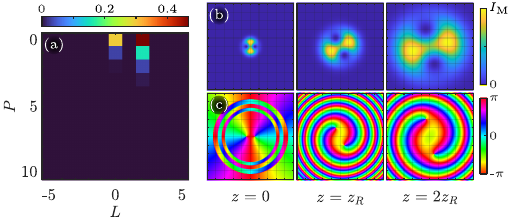}
    \caption{ Same as Fig. \ref{fig:Fig2} for two identical factors with $\{ p_{1}, \ell_{1} \} = \{ p_{2}, \ell_{2} \} = \{ 0, 1 \}$.}
    \label{fig:Fig4}
\end{figure}

We now consider pdLGBs that exhibit rigid rotation under free-space propagation.
This behavior arises when the factors have zero radial index $p_{j}=0$; share an identical azimuthal index $\ell_{j}=\ell$; and are placed evenly on a circle centered at the origin, $(x_{j},y_{j}) = r_{0} \left( \cos \left[ 2 \pi (j-1) / n + \theta_{0} \right], \sin \left[ 2 \pi (j-1) / n + \theta_{0} \right] \right)$, with radius $r_{0}$ and orientation $\theta_{0}$.
In this case, the transverse pattern rotates as a whole without deformation as the displaced coefficients share the same weight structure and accumulate a common Gouy phase.

\begin{figure}[ht]
    \centering
    \includegraphics[width =  0.75 \linewidth]{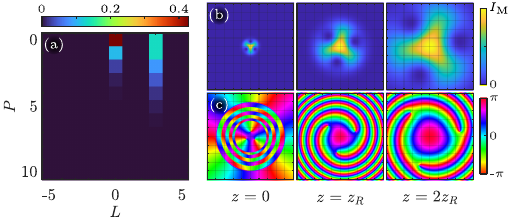}
    \caption{ Same as Fig. \ref{fig:Fig2} for three identical factors with $\{ p_{1}, \ell_{1} \} = \{ p_{2}, \ell_{2} \} = \{ p_{3}, \ell_{3} \} = \{ 0, 1 \}$ located at $\{ (x_{1},y_{1}), (x_{2},y_{2}), (x_{3},y_{3}) \} = \{ (w_{0}/2,0), (-w_{0}/4,\sqrt{3}w_{0}/4), (-w_{0}/4,-\sqrt{3}w_{0}/4) \}$.
    }
    \label{fig:Fig5}
\end{figure}

Figures~\ref{fig:Fig4} and \ref{fig:Fig5} illustrate rigid rotation for pdLGBs formed by identical displaced factors with $\{ p_{j}, \ell_{j} \} = \{ 0, 1 \}$ placed evenly on a circle.
Figure~\ref{fig:Fig4} shows two factors placed at diametrically opposite points. 
This configuration yields a net OAM of $\langle \hat{L}_{z} \rangle = 4\hbar/3$. 
Figure~\ref{fig:Fig5} extends the configuration to three factors at the vertices of an equilateral triangle, with $\langle \hat{L}_{z} \rangle = 1.28\hbar$.
In both cases, the modal expansion is restricted to a limited set of azimuthal indices, Fig.~\ref{fig:Fig4}(a) and Fig.~\ref{fig:Fig5}(a).
The intensity and phase distributions [Figs.~\ref{fig:Fig4}(b) and \ref{fig:Fig4}(c) and Figs.~\ref{fig:Fig5}(b) and \ref{fig:Fig5}(c)] rotate uniformly under propagation, while the transverse shape and relative geometry remain fixed, confirming rigid rotation.

\subsection{Nonrigid rotation}

\begin{figure}[ht]
    \centering
    \includegraphics[width =  0.75 \linewidth]{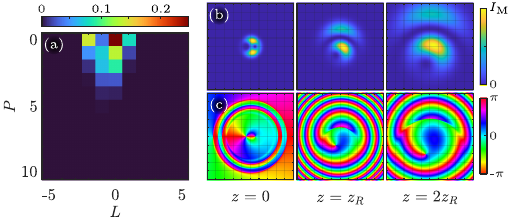}
    \caption{Same as Fig.~\ref{fig:Fig2} for two displaced LGBs with identical radial indices and unequal azimuthal charges,
    $\{ p_{1}, \ell_{1} \} = \{ 0, 1 \}$ and $\{ p_{2}, \ell_{2} \} = \{ 0, -2 \}$.
    }
    \label{fig:Fig6}
\end{figure}

\begin{figure}[ht]
    \centering
    \includegraphics[width =  0.75 \linewidth]{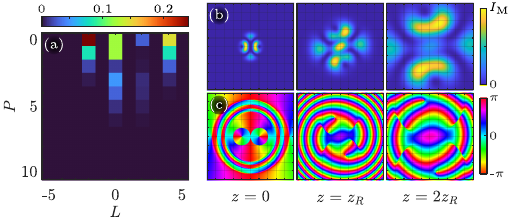}
    \caption{Same as Fig.~\ref{fig:Fig2} for two displaced LGBs with identical azimuthal indices and higher radial order,
    $\{ p_{1}, \ell_{1} \} = \{ p_{2}, \ell_{2} \} = \{ 1, 1 \}$, located at $\{ (x_{1},y_{1}), (x_{2},y_{2}) \} = \{(w_{0}, 0), (-w_{0}, 0)\}$.
   }
    \label{fig:Fig7}
\end{figure}

We finally consider pdLGBs that undergo nonrigid rotation under free-space propagation.
This regime arises when we relax the conditions for rigid rotation, either through unequal azimuthal indices or the presence of radial structure.
In these cases, the modal expansion distributes weight over multiple azimuthal sectors with different relative amplitudes, which cannot be absorbed uniformly into the displaced coefficients.
The transverse pattern rotates while its internal intensity distribution deforms during propagation.
Figures~\ref{fig:Fig6} and \ref{fig:Fig7} illustrate nonrigid rotation for pdLGBs formed by two displaced LGB factors.

Figure~\ref{fig:Fig6} shows the product of two displaced LGB factors with identical radial indices and unequal azimuthal charges, $\{ p_{1}, \ell_{1} \} = \{ 0, 1 \}$ and $\{ p_{2}, \ell_{2} \} = \{ 0, -2 \}$.
This configuration yields a net OAM of $\langle \hat{L}_{z} \rangle = -\hbar/2$.
In this case, the modal weights distribute asymmetrically over the azimuthal index, with contributions at $L \in \{-2,-1,0,1\}$ [Fig.~\ref{fig:Fig6}(a)].
The intensity and phase distributions [Figs.~\ref{fig:Fig6}(b) and \ref{fig:Fig6}(c)] rotate under propagation while the transverse structure deforms, breaking rigid rotation.

Figure~\ref{fig:Fig7} shows a different mechanism leading to nonrigid rotation, where both factors share identical azimuthal indices but have nonzero radial order, $\{ p_{1}, \ell_{1} \} = \{ p_{2}, \ell_{2} \} = \{ 1, 1 \}$.
This configuration yields a net OAM of $\langle \hat{L}_{z} \rangle = 0.45\hbar$.
Here, the modal weights are again asymmetric and sparsely distributed [Fig.~\ref{fig:Fig7}(a)] with dominant contributions at $L \in \{-2,0,2,4\}$.
Although the field rotates under propagation, the presence of multiple radial contributions with different weights leads to deformation of the transverse structure and redistribution of intensity [Figs.~\ref{fig:Fig7}(b) and \ref{fig:Fig7}(c)].

\section{Conclusions}
\label{sec:Sec4}
We developed an analytical framework for the free-space propagation of products of displaced Laguerre--Gaussian beams.

Our approach rewrites each displaced factor as an exact superposition of standard Laguerre--Gaussian beams centered at the origin, then collects the product into a single modal expansion with explicit complex coefficients providing a selection rule for azimuthal indices.
This turns propagation into the independent evolution of weighted standard Laguerre--Gaussian beam modes and makes the roles of curvature, transverse scaling, and Gouy phases transparent.

Working in the reference frame defined by the centroid of the transverse displacements aligns the propagation axis with the net orbital angular momentum of the field.
In this frame, the net orbital angular momentum follows directly from the modal weights of the expansion and provides a criterion to predict transverse rotation without numerical propagation.

Within our framework, we identify three distinct free-space propagation regimes.
First, products with vanishing net orbital angular momentum exhibit no transverse rotation, even though their intensity and phase distributions may redistribute during propagation.
Second, for identical displaced factors with zero radial index, modal weights and Gouy phases can be absorbed uniformly into the displacement coefficients, producing rigid rotation of the entire transverse structure.
Third, when we relax these conditions, either through unequal azimuthal indices or the presence of nonzero radial structure, this uniform redistribution is no longer possible, leading to nonrigid rotation accompanied by shifts in the transverse intensity and phase distributions.

Our framework provides a systematic path to engineer transverse rotation dynamics in structured light beams.
It complements existing approaches based on engineered point-spread functions and identifies products of displaced Laguerre--Gaussian beams as a platform for encoding propagation information into transverse rotation, with potential applications in depth-sensitive optical microscopy, imaging, and tracking.

\begin{acknowledgments}
B. M. R. L. acknowledges support and hospitality as an affiliate visiting colleague at the Department of Physics and Astronomy, University of New Mexico.
\end{acknowledgments}


%

\end{document}